\documentclass[rmp,titlepage,hidelinks,10pt,byrevtex,titlepage,secnumarabic,nofootinbib,longbibliography,a4paper,showkeys,preprintnumbers,twocolumn]{revtex4-2}
\usepackage{graphicx}
\usepackage{amsmath}
\usepackage{amssymb}
\usepackage{float}
\usepackage{moreverb}
\usepackage[commandnameprefix=always,final]{changes}

\usepackage{url}
\usepackage{svg}
\usepackage{tabularx,array}
\usepackage{ccicons}

\newcommand{\zdoi}{21641098}

\newcommand{\zenodo}[1]{\doi{10.5281/zenodo.#1}}

\usepackage{xr-hyper}
\usepackage{xcolor}
\usepackage{colortbl}
\usepackage[pdftex,
pdfauthor={Jose Maria Martin-Olalla and Jorge Mira},
pdftitle={Two Seconds per Day: Limitations of Circadian Burden Metrics in the Permanent Daylight Saving Time Debate},
            pdfsubject={DST and latitude, circadian disruption},
            pdfkeywords={DST; summer time; latitude; sleep deprivation; spring transition; Europe; insolation; season;motor vehicle accidents},
            pdfproducer={Latex with hyperref, or other system},
            pdfcreator={pdflatex, or other tool}]{hyperref}

            \newcommand\rorlink[2]{\href{https://ror.org/#2}{#1\,\includegraphics[scale=0.1]{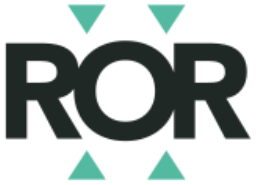}}
}
\usepackage{orcidlink}
\providecommand{\orcidlinki}[2]{\href{https://orcid.org/#2}{#1}\orcidlink{#2}}
            \usepackage{breakurl}
            \usepackage{siunitx}
            \DeclareSIUnit{\uyear}{a}
\usepackage{mathtools}
\usepackage[utf8]{inputenc}
\usepackage[english]{babel}
\usepackage{lineno}
\usepackage{array}

\AtBeginDocument{\usepackage{booktabs}}

\DeclareSIUnit{\year}{a}
\makeatletter\AtBeginDocument{\let\@elt\relax}\makeatother

\newcommand{\aucontribute}{\section*{Author contributions}}
\newcommand{\competing}{\section*{Declaration of interest}}
\newcommand{\funding}{\section*{Funding statement}}

\begin{document}
\author{\orcidlinki{José María Martín-Olalla}{0000-0002-3750-9113}}
\affiliation{\rorlink{Universidad de Sevilla}{03yxnpp24}, Facultad de Física, Departamento de Física de la Materia Condensada, ES41012 Sevilla, Spain}
\email{olalla@us.es}

\preprint{\textcolor{blue}{This is Authors' accepted manuscript posted on \zenodo{\zdoi}. The Version of Record \copyright Taylor and Francis 2026}}
  \preprint{\textcolor{blue}{ is available in \emph{Chronobiology International}  \doi{10.1080/07420528.2026.2745645}}}
\author{\orcidlinki{Jorge Mira}{0000-0002-6024-6294}}
\affiliation{\rorlink{Universidade de Santiago de Compostela}{030eybx10}, Facultade de Física, Departamento de Física Aplicada and iMATUS, ES15782 Santiago de Compostela, Spain}
\email{jorge.mira@usc.es}
\received[Submitted: ]{July 28th, 2026}
\accepted{October 2nd, 2026}
\published{October 15th, 2026}

\thanks{\ccby}
\title{Two Seconds per Day: Limitations of Circadian Burden Metrics in the Permanent Daylight Saving Time Debate}
  

\begin{abstract}

  A 1410-word Letter to the Editor reporting a bug in a study published in PNAS. The authors of the original study executed their code to compute the yearly circadian burden with an Earth rotating backwards: the sun rising in the West, setting on the East.  The authors have acknowledged the error and re-executed the code. We analyze authors's new average yearly circadian burden in the US to show that these results do not provide significant information for the current debate on the time policy.

    The Version of Record \copyright Taylor and Francis 2026 is published in \emph{Chronobiology International}   \doi{10.1080/07420528.2026.2745645}.

  This letter comments the study \doi{10.1073/pnas.2508293122}.

  This letter continues the letter \doi{10.1073/pnas.2532075123}.

  This letter is supplemented by the preprint \doi{10.64898/2026.06.17.26355889}.
  
  Authors' reply can be found at \doi{10.1080/07420528.2026.2744598}.

  This is posted in \zenodo{\zdoi}.

  \textcolor{blue}{Existe una versión en lengua castellana.}

  Word count: 1410

\end{abstract}

\maketitle
The logic underlying seasonal clock adjustments is ostensibly straightforward: it is a synchronized administrative mechanism designed to align morning social schedules with the seasonal oscillations of sunrise triggered by Earth’s axial tilt at extratropical latitudes \cite{Hudson1895}. Yet, this apparent simplicity evaporates when modeling the complex interplay between immutable solar cycles and plastic human behavior. Clock regulations highlight a fundamental question: whether it is rational to advance social schedules in spring and summer in response to earlier sunrises. 

The natural experiment created by biannual clock changes has led to a myriad of studies reporting minor acute effects associated with transition dates, chiefly the spring transition to Daylight Saving Time (DST) \cite{Janszky2009,Fritz2020,Culic2013}. Here, science successfully isolates acute transient effects from background seasonal variation. Another set of natural experiments examines time-zone boundaries, reporting a slight degradation of health metrics on the eastern side of a time-zone border \cite{Giuntella2019,Gentry2022}, suggesting that permanent DST (pDST) may carry greater health risks than permanent Standard Time (pST). However, a fundamental empirical gap remains: no study has demonstrated that living under DST during summer is inherently hazardous, because existing methodologies fail to decouple the administrative rationale of clock adjustments from pure, latitude-dependent seasonal photoperiod dynamics.\cite{Martin-Olalla2024g}

Recently, \citet{Weed2025} claimed that a mathematical circadian model could directly attribute disease prevalence in the United States to exposure to seasonal clocks. Their results indicated that current seasonal clock switching yields the highest annual circadian burden ($Y$), followed by locked pDST, with locked pST yielding the lowest burden. \citet{Erren2026} noted that the study appears susceptible to an ecological fallacy by cross-correlating individual-level circadian stressors with aggregate societal health outcomes. Furthermore, \citet{Martin-Olalla2026c} demonstrated that the metric $Y$ merely aggregates the absolute values of minute daily circadian shifts. The reported annual burden $Y$, which aggregates 364 daily absolute shifts, amounts to approximately \qty{20}{\hour} per person per year, or roughly \qty{3}{\minute} per person per day. This corresponds to a daily variation of less than \qty{0.2}{\percent} of the 24-hour circadian period, a magnitude far too small to establish a plausible biological mechanism linking it to chronic disease prevalence. Additionally, while the original source code \cite{Weed2025b} was subject to a published correction \cite{Weed2026b}, the core findings were claimed to hold.

Here, we report another critical implementation error in the original codebase that fundamentally impacts these conclusions. \citet[Figure~2c]{Weed2025} reported the ``hours from time zone center'', also known as the longitudinal time offset ($t_o$), defined as the difference between local solar noon at a county centroid and solar noon at the time zone's reference meridian. In their spatial visualization, negative values of $t_o$ (shading orange) represent locations west of the reference meridian, while positive values (shading yellow) represent locations to the east. Administrative seasonal clock shifts effectively modify $t_o$: advancing the clock by \qty{15}{\degree} east in spring decreases the offset ($\Delta t_o = \qty{-1}{\hour}$), whereas setting it back in autumn increases the offset ($\Delta t_o = \qty{1}{\hour}$).

In computing the circadian model, the original code assigned Earth's rotational speed a value of \qty{-15}{\degree\per\hour} instead of \qty{15}{\degree\per\hour}. This error effectively modeled an inverted Earth spinning westward, with the sun rising in the west and setting in the east. \chreplaced{Alternatively, the bug}{Geographically, this} flipped U.S. counties across their East--West axes, placing, for instance, Amarillo, Texas, east of the eastern boundary of the Central Time Zone, and shifting Detroit, Michigan, east of the Eastern Time Zone edge into the middle of the Atlantic Ocean (for further technical details, see \citet{Martin-Olalla2026d}). \chadded{Consequently, the model evaluated circadian metrics calculated for these geographically fictitious coordinates against empirical health prevalence data from their real-world counterparts.}

Following our communication of this error, Weed re-ran the model and communicated updated population-weighted annual circadian burdens\chadded{, and a draft heat map replicating \citet[Figure~3]{Weed2025}}. She reported that the results ``continue to show the lowest burden under permanent Standard Time and the highest burden under a biannual switching policy.'' \chreplaced{Her}{The} updated metrics shifted as follows: pST moved from \qty{18.5}{\hour} to \qty{19.2}{\hour} per person-year; pDST moved from \qty{19.6}{\hour} to \qty{19.4}{\hour} per person-year; and seasonal clocks moved from \qty{20.9}{\hour} to \qty{21.0}{\hour} per person-year (Table~\ref{tab:to}).

\chreplaced{Here, we analyze Weed's corrected average values and reveal a series of critical inconsistencies:}{These corrected numbers reveal several critical inconsistencies:}

First, the updated difference between $Y_{\text{pST}}$ and $Y_{\text{pDST}}$ is now just \qty{0.2}{\hour} per person-year (\qty{12}{\minute} per person-year), which translates to an average daily difference of approximately \qty{2}{\second} per person. \chadded{The convergence of the pST and pDST new solutions is equally evident in the draft heat-map provided by Weed.} Claiming that pST is biologically superior to pDST based on a 2-second daily variance is scientifically untenable; within the numerical resolution of the model, pST and pDST yield functionally identical outcomes. Furthermore, if $Y$ were indeed a meaningful descriptor of health hazards, it would be impossible to reconcile this near-zero policy difference with empirical observations at time-zone boundaries \cite{Giuntella2019,Gentry2022}, which consistently document tangible health and economic disparities between the eastern and western edges of a time zone.

Second, relying primarily on aggregate average values exposes a key methodological vulnerability: the study does not rigorously analyze county-level associations between $Y$ and disease prevalence, an analysis that would have been heavily distorted by spatial inversion across individual counties. Instead, it relies on aggregate outputs for average $Y$. Because $Y$ is a smooth, continuous function of latitude ($\phi$) and longitudinal offset ($t_o$), the population-weighted average burden across all US counties is closely approximated by evaluating $Y$ directly at the population-weighted mean coordinates: $\langle\phi\rangle = \qty{37.4}{\degree}$ and $\langle t_o\rangle = \qty{-12}{\minute}$. The code's East–West inversion mistakenly shifted this mean time offset from \qty{-12}{\minute} to \qty{+12}{\minute}, introducing a net \qty{24}{\minute} spatial discrepancy ($\Delta t_o = \qty{24}{\minute}$).

Nonetheless, by comparing the uncorrected and corrected outputs, we can evaluate the model's sensitivity ($\Delta Y / \Delta t_o$) to longitudinal placement. For pST, the East–West correction increased the burden by \qty{0.7}{\hour} per year ($\qty{19.2}{\hour} - \qty{18.5}{\hour}$), yielding a gradient of \qty{1.8}{\hour} per year per hour of offset shift. For pDST, the correction decreased the burden by \qty{0.2}{\hour} per year, yielding a negative gradient of \qty{-0.5}{\hour} per year per hour of offset shift. 

Furthermore, comparing pST against pDST under the uncorrected code implied a policy gradient of \qty{1.1}{\hour} per year per hour of offset shift ($\qty{19.6}{\hour} - \qty{18.5}{\hour}$). Under the corrected code, this policy gradient collapses to just \qty{0.2}{\hour} per year per hour of offset shift ($\qty{19.4}{\hour} - \qty{19.2}{\hour}$), a 5.5-fold reduction in sensitivity (Table~\ref{tab:to}). The lack of monotonicity and stability across these directional inversions demonstrates that the model's output is governed by numerical instability and background noise rather than robust circadian dynamics. We suspect that the 5-minute integration time step ($\delta = \qty{5}{\minute}$) used in the baseline calculations is too coarse to provide stable spatial resolution.

\begin{table*}[t]
\centering
  \begin{tabular}{lrrl}
    \toprule
    \textbf{Scenario} & $t_o$ & $Y$ (\qty{}{\hour\per\year}) & \textbf{Description} \\
    \midrule
    $A$ & \qty{+12}{\minute} & 18.5 & pST flipped distribution (original/uncorrected) \\
    $B$ & \qty{-48}{\minute} & 19.6 & pDST flipped distribution (original/uncorrected) \\
    $C$ & \qty{-12}{\minute} & 19.2 & pST correct distribution (updated) \\
    $D$ & \qty{-72}{\minute} & 19.4 & pDST correct distribution (updated) \\
    \midrule
    \multicolumn{4}{c}{\textbf{Calculated Gradients} ($\Delta Y / \Delta t_o$ in \qty{}{\hour\per\hour})} \\
    \midrule
    $B - A$ & & \qty{1.1}{\hour\per\hour} & pST to pDST shift (uncorrected code) \\
    $D - C$ & & \qty{0.2}{\hour\per\hour} & pST to pDST shift (corrected code) \\
    $C - A$ & & \qty{1.8}{\hour\per\hour} & East--West inversion effect on pST \\
    $D - B$ & & \qty{-0.5}{\hour\per\hour} & East--West inversion effect on pDST \\
    \bottomrule
  \end{tabular}
  \caption{Top: \chreplaced{Weed's reported}{Reported} values of the annual population-weighted circadian burden $Y$ \chdeleted{per person-year} under different scenarios defined by the weighted mean time offset $t_o$. Bottom: \chreplaced{Our derived}{Derived} gradients computed from \chadded{Weed's} reported outputs. The lack of numerical stability and monotonicity indicates that the model metrics are heavily influenced by resolution artifacts.}
  \label{tab:to}
\end{table*}

Finally, the remaining \qty{2}{\hour} annual burden gap between seasonal clock switching (\qty{21.0}{\hour}) and locked clock policies (\qty{19.2}{\hour}--\qty{19.4}{\hour}) matches the exact cumulative sum of the two mandatory 1-hour step changes introduced during the spring and autumn transitions (\qty{1}{\hour} + \qty{1}{\hour} = \qty{2}{\hour}). This indicates that the reported ``excess burden'' under seasonal clocks is simply a direct mathematical tally of the discrete clock shifts themselves, rather than evidence of sustained, long-term physiological misalignment. \chadded{A true physiological interaction with the seasonal photoperiod would yield an excess burden distinct from this \qty{2}{\hour} baseline, either larger (if transitions compound misalignment) or smaller (if they alleviate it). An exact match of \qty{2}{\hour} offers no evidence of sustained physiological strain beyond the transient step changes. Coupled with the virtual identity $Y_{\mathrm{pST}} \approx Y_{\mathrm{pDST}}$, this underscores the lack of policy-relevant physiological signal in these metrics.}

In conclusion, the corrected metrics reported by Weed fail to support the primary conclusions of the original study. Specifically: (1) pST displays no meaningful biological advantage over pDST, and (2) the model fails to isolate chronic exposure to seasonal time policies from basic transition artifacts and spatial resolution noise.

\begin{acknowledgments}
  
The authors acknowledge the use of an Artificial Intelligence large language model, Gemini, to refine the draft manuscript for grammar and clarity. Gemini was made available to the authors through a collaborative initiative between Google and Universidad de Sevilla.
\end{acknowledgments}

\aucontribute

All authors contributed equally to this work.

\competing

The authors declare no competing interest.

\funding

This work was not funded.

%

\section*{Manuscript's time line}
After the publishing of our Letter\cite{Martin-Olalla2026c} and reading authors' subsequent reply\cite{Weed2026}, the curiosity bug bit us. We decided to pull down and run the original code ourselves to compute and observe the yearly burden.

The initial execution for the \num{3108} counties took several days, which gave us ample time to dive deep into the underlying methodology. That's when we noticed two critical issues. First, the methodology prioritized clock time over solar time, which is often a really bad take. 

Second, as we looked closer, all of a sudden, we uncovered the inversion error. It, too, was sitting right there in plain sight, but we had initially trusted the comments embedded withtin the script rather than checking the raw mechanics. We checked that the light diets were inverted: Western locations had brighter mornings and darker evenings in their clock time analysis. The issue was hidding at plain sight, but we must admit that when writing our initial Letter we had trusted Authors' framing and did not scrutinize their methodology thoroughly.

 The writing of the manuscript required \texttt{66
} different \texttt{emacs} sessions from \texttt{2026-04-16 18:36
} to \mbox{\texttt
{2026-10-16 17:54
}.}


\begin{thebibliography}{14}%
\makeatletter
\providecommand \@ifxundefined [1]{%
 \@ifx{#1\undefined}
}%
\providecommand \@ifnum [1]{%
 \ifnum #1\expandafter \@firstoftwo
 \else \expandafter \@secondoftwo
 \fi
}%
\providecommand \@ifx [1]{%
 \ifx #1\expandafter \@firstoftwo
 \else \expandafter \@secondoftwo
 \fi
}%
\providecommand \natexlab [1]{#1}%
\providecommand \enquote  [1]{``#1''}%
\providecommand \bibnamefont  [1]{#1}%
\providecommand \bibfnamefont [1]{#1}%
\providecommand \citenamefont [1]{#1}%
\providecommand \href@noop [0]{\@secondoftwo}%
\providecommand \href [0]{\begingroup \@sanitize@url \@href}%
\providecommand \@href[1]{\@@startlink{#1}\@@href}%
\providecommand \@@href[1]{\endgroup#1\@@endlink}%
\providecommand \@sanitize@url [0]{\catcode `\\12\catcode `\$12\catcode
  `\&12\catcode `\#12\catcode `\^12\catcode `\_12\catcode `\%12\relax}%
\providecommand \@@startlink[1]{}%
\providecommand \@@endlink[0]{}%
\providecommand \url  [0]{\begingroup\@sanitize@url \@url }%
\providecommand \@url [1]{\endgroup\@href {#1}{\urlprefix }}%
\providecommand \urlprefix  [0]{URL }%
\providecommand \Eprint [0]{\href }%
\providecommand \doibase [0]{https://doi.org/}%
\providecommand \selectlanguage [0]{\@gobble}%
\providecommand \bibinfo  [0]{\@secondoftwo}%
\providecommand \bibfield  [0]{\@secondoftwo}%
\providecommand \translation [1]{[#1]}%
\providecommand \BibitemOpen [0]{}%
\providecommand \bibitemStop [0]{}%
\providecommand \bibitemNoStop [0]{.\EOS\space}%
\providecommand \EOS [0]{\spacefactor3000\relax}%
\providecommand \BibitemShut  [1]{\csname bibitem#1\endcsname}%
\let\auto@bib@innerbib\@empty
\bibitem [{\citenamefont {Anonymous}(2026)}]{Weed2026b}%
  \BibitemOpen
  \bibfield  {author} {\bibinfo {author} {\bibnamefont {Anonymous},}} (\bibinfo
  {year} {2026}),\ \bibfield  {title} {\enquote {\bibinfo {title} {Correction
  for weed and zeitzer, circadian-informed modeling predicts regional variation
  in obesity and stroke outcomes under different permanent us time policies},}\
  }\href {https://doi.org/10.1073/PNAS.2612351123} {\bibfield  {journal}
  {\bibinfo  {journal} {Proceedings of the National Academy of Sciences}\
  }\textbf {\bibinfo {volume} {123}},\ \bibinfo {pages}
  {e2612351123}}\BibitemShut {NoStop}%
\bibitem [{\citenamefont {Erren}\ \emph {et~al.}(2026)\citenamefont {Erren},
  \citenamefont {Lewis},\ and\ \citenamefont {Morfeld}}]{Erren2026}%
  \BibitemOpen
  \bibfield  {author} {\bibinfo {author} {\bibnamefont {Erren}, \bibfnamefont
  {Thomas~C}}, \bibinfo {author} {\bibfnamefont {Philip}\ \bibnamefont
  {Lewis}}, and\ \bibinfo {author} {\bibfnamefont {Peter}\ \bibnamefont
  {Morfeld}}} (\bibinfo {year} {2026}),\ \bibfield  {title} {\enquote {\bibinfo
  {title} {Challenges of county-level circadian modeling for time-policy
  debates},}\ }\href {https://doi.org/10.1073/PNAS.2527389123} {\bibfield
  {journal} {\bibinfo  {journal} {Proceedings of the National Academy of
  Sciences}\ }\textbf {\bibinfo {volume} {123}},\ \bibinfo {pages}
  {e2527389123}}\BibitemShut {NoStop}%
\bibitem [{\citenamefont {Fritz}\ \emph {et~al.}(2020)\citenamefont {Fritz},
  \citenamefont {VoPham}, \citenamefont {Wright},\ and\ \citenamefont
  {Vetter}}]{Fritz2020}%
  \BibitemOpen
  \bibfield  {author} {\bibinfo {author} {\bibnamefont {Fritz}, \bibfnamefont
  {Josef}}, \bibinfo {author} {\bibfnamefont {Trang}\ \bibnamefont {VoPham}},
  \bibinfo {author} {\bibfnamefont {Kenneth~P.}\ \bibnamefont {Wright}}, and\
  \bibinfo {author} {\bibfnamefont {Céline}\ \bibnamefont {Vetter}}} (\bibinfo
  {year} {2020}),\ \bibfield  {title} {\enquote {\bibinfo {title} {A
  chronobiological evaluation of the acute effects of daylight saving time on
  traffic accident risk},}\ }\href {https://doi.org/10.1016/j.cub.2019.12.045}
  {\bibfield  {journal} {\bibinfo  {journal} {Current Biology}\ }\textbf
  {\bibinfo {volume} {30}},\ \bibinfo {pages} {729--735}}\BibitemShut {NoStop}%
\bibitem [{\citenamefont {Gentry}\ \emph {et~al.}(2022)\citenamefont {Gentry},
  \citenamefont {Evaniuck}, \citenamefont {Suriyamongkol},\ and\ \citenamefont
  {Mali}}]{Gentry2022}%
  \BibitemOpen
  \bibfield  {author} {\bibinfo {author} {\bibnamefont {Gentry}, \bibfnamefont
  {Jeffery}}, \bibinfo {author} {\bibfnamefont {Jayson}\ \bibnamefont
  {Evaniuck}}, \bibinfo {author} {\bibfnamefont {Thanchira}\ \bibnamefont
  {Suriyamongkol}}, and\ \bibinfo {author} {\bibfnamefont {Ivana}\ \bibnamefont
  {Mali}}} (\bibinfo {year} {2022}),\ \bibfield  {title} {\enquote {\bibinfo
  {title} {Living in the wrong time zone: Elevated risk of traffic fatalities
  in eccentric time localities},}\ }\href
  {https://doi.org/10.1177/0961463x221104675} {\bibfield  {journal} {\bibinfo
  {journal} {Time and Society}\ }\textbf {\bibinfo {volume} {31}},\ \bibinfo
  {pages} {457--479}}\BibitemShut {NoStop}%
\bibitem [{\citenamefont {Giuntella}\ and\ \citenamefont
  {Mazzonna}(2019)}]{Giuntella2019}%
  \BibitemOpen
  \bibfield  {author} {\bibinfo {author} {\bibnamefont {Giuntella},
  \bibfnamefont {Osea}}, and\ \bibinfo {author} {\bibfnamefont {Fabrizio}\
  \bibnamefont {Mazzonna}}} (\bibinfo {year} {2019}),\ \bibfield  {title}
  {\enquote {\bibinfo {title} {Sunset time and the economic effects of social
  jetlag: evidence from us time zone borders},}\ }\href
  {https://doi.org/10.1016/j.jhealeco.2019.03.007} {\bibfield  {journal}
  {\bibinfo  {journal} {Journal of Health Economics}\ }\textbf {\bibinfo
  {volume} {65}},\ \bibinfo {pages} {210--226}}\BibitemShut {NoStop}%
\bibitem [{\citenamefont {Hudson}(1895)}]{Hudson1895}%
  \BibitemOpen
  \bibfield  {author} {\bibinfo {author} {\bibnamefont {Hudson}, \bibfnamefont
  {George~Vernon}}} (\bibinfo {year} {1895}),\ \bibfield  {title} {\enquote
  {\bibinfo {title} {On seasonal time-adjustment in countries south of lat. 30
  degrees},}\ }\href
  {https://paperspast.natlib.govt.nz/periodicals/TPRSNZ1895-28.2.7.1.5}
  {\bibfield  {journal} {\bibinfo  {journal} {Transactions and Proceedings of
  the Royal Society of New Zealand}\ }\textbf {\bibinfo {volume} {28}},\
  \bibinfo {pages} {734}}\BibitemShut {NoStop}%
\bibitem [{\citenamefont {Janszky}\ and\ \citenamefont
  {Ljung}(2008)}]{Janszky2009}%
  \BibitemOpen
  \bibfield  {author} {\bibinfo {author} {\bibnamefont {Janszky}, \bibfnamefont
  {Imre}}, and\ \bibinfo {author} {\bibfnamefont {Rickard}\ \bibnamefont
  {Ljung}}} (\bibinfo {year} {2008}),\ \bibfield  {title} {\enquote {\bibinfo
  {title} {Shifts to and from daylight saving time and incidence of myocardial
  infarction},}\ }\href {https://doi.org/10.1056/NEJMC0807104} {\bibfield
  {journal} {\bibinfo  {journal} {New England Journal of Medicine}\ }\textbf
  {\bibinfo {volume} {359}},\ \bibinfo {pages} {1966--1968}}\BibitemShut
  {NoStop}%
\bibitem [{\citenamefont {Martín-Olalla}\ and\ \citenamefont
  {Mira}(2025)}]{Martin-Olalla2024g}%
  \BibitemOpen
  \bibfield  {author} {\bibinfo {author} {\bibnamefont {Martín-Olalla},
  \bibfnamefont {José~María}}, and\ \bibinfo {author} {\bibfnamefont {Jorge}\
  \bibnamefont {Mira}}} (\bibinfo {year} {2025}),\ \bibfield  {title} {\enquote
  {\bibinfo {title} {Assessing the best hour to start the day: an appraisal of
  seasonal daylight saving time},}\ }\href
  {https://doi.org/10.1098/rsos.240727} {\bibfield  {journal} {\bibinfo
  {journal} {The Royal Society Open Science}\ }\textbf {\bibinfo {volume}
  {12}},\ \bibinfo {pages} {240727}}\BibitemShut {NoStop}%
\bibitem [{\citenamefont {Martín-Olalla}\ and\ \citenamefont
  {Mira}(2026{\natexlab{a}})}]{Martin-Olalla2026d}%
  \BibitemOpen
  \bibfield  {author} {\bibinfo {author} {\bibnamefont {Martín-Olalla},
  \bibfnamefont {José~María}}, and\ \bibinfo {author} {\bibfnamefont {Jorge}\
  \bibnamefont {Mira}}} (\bibinfo {year} {2026}{\natexlab{a}}),\ \bibfield
  {title} {\enquote {\bibinfo {title} {Methodological guidelines for circadian
  modeling of daylight saving time: application to the united states},}\ }\href
  {https://doi.org/10.64898/2026.06.17.26355889} {\bibinfo  {journal}
  {medRxiv}\ ,\ \bibinfo {pages} {2026.06.17.26355889}}\BibitemShut {NoStop}%
\bibitem [{\citenamefont {Martín-Olalla}\ and\ \citenamefont
  {Mira}(2026{\natexlab{b}})}]{Martin-Olalla2026c}%
  \BibitemOpen
\bibfield  {journal} {  }\bibfield  {author} {\bibinfo {author} {\bibnamefont
  {Martín-Olalla}, \bibfnamefont {José~María}}, and\ \bibinfo {author}
  {\bibfnamefont {Jorge}\ \bibnamefont {Mira}}} (\bibinfo {year}
  {2026}{\natexlab{b}}),\ \bibfield  {title} {\enquote {\bibinfo {title} {The
  sum of absolute circadian shifts: questioning the metric linking daylight
  saving time policy to health issues.}}\ }\href
  {https://doi.org/10.1073/PNAS.2532075123} {\bibfield  {journal} {\bibinfo
  {journal} {Proceedings of the National Academy of Sciences}\ }\textbf
  {\bibinfo {volume} {123}},\ \bibinfo {pages} {e2532075123}}\BibitemShut
  {NoStop}%
\bibitem [{\citenamefont {Weed}(2025)}]{Weed2025b}%
  \BibitemOpen
  \bibfield  {author} {\bibinfo {author} {\bibnamefont {Weed}, \bibfnamefont
  {Lara}}} (\bibinfo {year} {2025}),\ \href
  {https://github.com/Lara-Weed/CircadianTimePolicy} {\emph {\bibinfo {title}
  {Circadian Time Policy}}},\ \bibinfo {type} {Tech. Rep.}\ (\bibinfo
  {institution} {GitHub})\BibitemShut {NoStop}%
\bibitem [{\citenamefont {Weed}\ and\ \citenamefont
  {Zeitzer}(2025)}]{Weed2025}%
  \BibitemOpen
  \bibfield  {author} {\bibinfo {author} {\bibnamefont {Weed}, \bibfnamefont
  {Lara}}, and\ \bibinfo {author} {\bibfnamefont {Jamie~M.}\ \bibnamefont
  {Zeitzer}}} (\bibinfo {year} {2025}),\ \bibfield  {title} {\enquote {\bibinfo
  {title} {Circadian-informed modeling predicts regional variation in obesity
  and stroke outcomes under different permanent us time policies},}\ }\href
  {https://doi.org/10.1073/PNAS.2508293122} {\bibfield  {journal} {\bibinfo
  {journal} {Proceedings of the National Academy of Sciences of the United
  States of America}\ }\textbf {\bibinfo {volume} {122}},\ \bibinfo {pages}
  {e2508293122}}\BibitemShut {NoStop}%
\bibitem [{\citenamefont {Weed}\ and\ \citenamefont
  {Zeitzer}(2026)}]{Weed2026}%
  \BibitemOpen
  \bibfield  {author} {\bibinfo {author} {\bibnamefont {Weed}, \bibfnamefont
  {Lara}}, and\ \bibinfo {author} {\bibfnamefont {Jamie~M.}\ \bibnamefont
  {Zeitzer}}} (\bibinfo {year} {2026}),\ \bibfield  {title} {\enquote {\bibinfo
  {title} {Reply to erren et al. and martin-olalla and mira: What constitutes
  circadian burden? an empirically grounded approach to long-term summary
  metrics},}\ }\href {https://doi.org/10.1073/PNAS.2533444123} {\bibfield
  {journal} {\bibinfo  {journal} {Proceedings of the National Academy of
  Sciences}\ }\textbf {\bibinfo {volume} {123}},\ \bibinfo {pages}
  {e2533444123}}\BibitemShut {NoStop}%
\bibitem [{\citenamefont {Čulić}(2013)}]{Culic2013}%
  \BibitemOpen
  \bibfield  {author} {\bibinfo {author} {\bibnamefont {Čulić}, \bibfnamefont
  {Viktor}}} (\bibinfo {year} {2013}),\ \bibfield  {title} {\enquote {\bibinfo
  {title} {Daylight saving time transitions and acute myocardial infarction},}\
  }\href {https://doi.org/10.3109/07420528.2013.775144} {\bibfield  {journal}
  {\bibinfo  {journal} {Chronobiology International}\ }\textbf {\bibinfo
  {volume} {30}},\ \bibinfo {pages} {662--668}}\BibitemShut {NoStop}%
\end{thebibliography}
\end{document}